\documentclass[nofootinbib,a4paper,aps,prx, twocolumn,amsmath]{revtex4-2}
\usepackage{graphicx}
\usepackage{amsfonts}
\usepackage{color}
\usepackage[colorlinks=true,citecolor=blue,linkcolor=blue,urlcolor=blue]{hyperref}
\pdfoutput=1
\usepackage[utf8]{inputenc}
\usepackage[english]{babel}
\usepackage[T1]{fontenc}
\usepackage{amsmath}
\usepackage{amssymb}
\usepackage{hyperref}

\usepackage{tikz}
\usepackage{lipsum}

\usepackage{mathtools}
\usepackage{units}
\usepackage{graphicx}
\usepackage{bm}
\usepackage{dsfont}
\usepackage{ulem}
\usepackage{mathrsfs}
\usepackage{anysize}

\DeclarePairedDelimiter\ket{\lvert}{\rangle}
\DeclarePairedDelimiter\expval{\langle}{\rangle}
\DeclarePairedDelimiterX\braket[2]{\langle}{\rangle}{#1\,\delimsize\vert\,\mathopen{}#2}
\newcommand{\Op}{\hat{\mathcal{O}}}
\newcommand{\Ham}{\hat{\mathcal{H}}}
\newcommand{\Rhop}{\hat{\rho}}
\DeclarePairedDelimiter\Tr{\mathrm{Tr}\{}{\}}

\begin{document}


\title{Pauli weight requirement of the matrix elements in time-evolved local operators: dependence beyond the  equilibration temperature}

\author{Carlos Ramos-Marim\'on}
\affiliation{Departament de F\'{\i}sica Qu\`antica i Astrof\'{\i}sica and Institut de Ci\`encies del Cosmos (ICCUB), Universitat de Barcelona,  Mart\'{\i} i Franqu\`es 1, 08028 Barcelona, Catalonia, Spain}
\email{carlos.ramos.marimon@icc.ub.edu}
\author{Stefano Carignano}
\affiliation{Barcelona Supercomputing Center}
\email{stefano.carignano@bsc.es}
\author{Luca Tagliacozzo}
\affiliation{Institute of Fundamental Physics IFF-CSIC, Calle Serrano 113b, Madrid 28006, Spain}
\email{luca.tagliacozzo@iff.csic.es}
\begin{abstract}
The complexity of simulating the out-of-equilibrium evolution of local operators in the Heisenberg picture is governed by the operator entanglement, which grows linearly in time for generic non-integrable systems, leading to an exponential increase in computational resources. A promising approach to simplify this challenge involves discarding parts of the operator and focusing on a subspace formed by "light" Pauli strings—strings with few Pauli matrices—as proposed by \textit{Rakovszki et al.} [PRB 105, 075131 (2022)]. In this work, we investigate whether this strategy can be applied to quenches starting from homogeneous product states. For ergodic dynamics, these initial states grant access to a wide range of equilibration temperatures. By concentrating on the desired matrix elements and retaining only the portion of the operator that contains Pauli strings parallel to the initial state, we uncover a complex scenario. In some cases, the light Pauli strings suffice to describe the dynamics, enabling efficient simulation with current algorithms. However, in other cases, heavier strings become necessary, pushing computational demands beyond our current capabilities. We analyze this behavior using a newly introduced measure of complexity, the Operator Weight Entropy, which we compute for different operators across most points on the Bloch sphere.
\end{abstract}

\maketitle

\section{Introduction}

Understanding the out-of-equilibrium dynamics of many-body quantum systems remains one of the major challenges in modern physics \cite{eisert2010colloquium}. Being able to predict and control this dynamics is crucial for designing protocols to protect quantum information, generating strongly correlated states of matter, and testing fundamental hypotheses about quantum evolution. Such an out-of-equilibrium evolution is indeed considered the ideal scenario of quantum advantange \cite{preskill2012quantum, daley2022practical}, where quantum simulators or quantum computers can be used to provide information that is not accessible by performing classical computations. 

In this paper, we focus on out-of-equilibrium dynamics induced by quantum quenches in closed systems \cite{polkovnikov2011colloquium}, where the system is prepared in an initial state and driven out of equilibrium by a sudden change in the Hamiltonian. We study the time-dependence of the expectation values of local operators, which typically oscillate and eventually equilibrate to a stationary value.

For ergodic systems, the long-time stationary value should coincide with the thermal equilibrium value as predicted by the Eigenstate Thermalization Hypothesis (ETH) \cite{deutsch1991quantum, srednicki1994chaos, srednicki1999approach}, confirming that traditional statistical mechanics provides a correct description of the late-time behavior. However, numerous scenarios have emerged where ergodicity is broken—ranging from integrable dynamics \cite{rigol2008thermalization, vidmar2016generalized} to many-body localization \cite{schreiber2015observation, smith2016manybody, lukin2019probing}, as well as cases where specific initial states delay or evade thermalization \cite{turner2018quantum, bernien2017probing, ho2019periodic, serbyn2021quantum}, thereby violating ETH. These phenomena are still heavily debated due to the lack of reliable computational tools that can capture the late-time dynamics of strongly correlated quantum systems.

Tensor Network (TN) algorithms \cite{bridgeman2017handwaving, ran2020tensor} have emerged as privileged candidates to describe the dynamics, due to their success at least for short times \cite{vidal2003efficient, vidal2004efficient, white2004realtime, feiguin2005timestep, daley2004timedependent, paeckel2019timeevolution}. Current TN strategies evolving initial states (Schrödinger picture) or operators (Heisenberg picture) face significant challenges due to the exponential increase in complexity. In the Schrödinger picture, this manifests as a rapid growth of entanglement entropy \cite{calabrese2005evolution, chiara2006entanglementa, perales2008entanglement, lauchli2008spreading, kim2013ballistic, torlai2014dynamics}, while in the Heisenberg picture, the complexity arises from the linear growth of operator entanglement \cite{zanardi2001entanglement, prosen2007operator, znidaric2008manybody, pizorn2009operator, dubail2017entanglement, zhou2017operator}.

While the complexity of evolving states in the Schrödinger picture is somewhat understood—for example, small global quenches near the ground state energy are manageable using mean-field approaches—the complexity in the Heisenberg picture remains less clear. Notably, when evolving local operators, there is no analog to temperature that can classify quenches as "large" or "small". Instead, the relationship between the initial state and the quench Hamiltonian only enters at the end of the evolution. This implies that in the Heisenberg picture one has to first build the full time-evolved operator before extracting its possible simpler matrix elements relevant to the quench.

Partial understanding has been achieved through the concept of \textit{operator entanglement}, which quantifies the difficulty of representing an operator as a Matrix Product State (MPS). Integrable systems are known to generate less operator entanglement compared to chaotic systems, with the former showing at most logarithmic growth over time, while the latter typically exhibit a linear growth \cite{prosen2007operator, pizorn2009operator}. This reduction in complexity for integrable systems is often attributed to their closed operator algebras.

However, even when local operators thermalize, the dynamics typically needs to overcome an "entanglement barrier", a regime where complexity becomes overwhelming before later-time equilibration simplifies the dynamics. This concept is well-established in the Schrödinger picture, and in the Heisenberg picture a similar phenomenon happens in terms of operator entanglement \cite{dubail2017entanglement}. Recent works have proposed avenues to overcome such an "operator entanglement barrier" by projecting the Heisenberg operators into the subspace with a few non-trivial Pauli operators, that is on the subspace of light Pauli strings \cite{rakovszky2022dissipationassisted, lloyd2024ballistic, kuo2024energy}. This prescriptions were motivated by results drawn from the field of operator spreading in random circuits \cite{vonkeyserlingk2018operator, khemani2018operator, rakovszky2018diffusive}, and proved to yield accurate results on high-temperature transport coefficients for conserved quantities, but their generally applicability is still unclear.

In this work, we try to better characterize the origin of the "operator entanglement barrier" for the evolution of local operators in the Heisenberg picture \cite{dubail2017entanglement}. We focus on quantum quenches from homogeneous Product States (PS) in a one-dimensional spin-1/2 chain governed by the Ising Hamiltonian with both parallel and transverse fields. In order to analyze the complexity of the operator evolution at a finer level than what we obtain from the simple operator entanglement, we introduce the concept of  \textit{Operator Weight Entropy} (OWE), which quantifies the complexity of the evolved operator  in terms of its content of non-trivial Pauli strings, that is by classifying the strings by their weights. We discuss the relation between heavy and light Pauli strings and the operator entanglement spectrum that dictate the operator entanglement entropy. We find that the two are not always related, and in some cases heavy operators can have support on the leading part of the operator entanglement spectrum, while in other cases they are supported on the tails. 

By examining various local operators, we identify an \textit{OWE barrier}—an intermediate regime where the operator complexity is high, analogous to the entanglement barrier in the Schrödinger picture. Interestingly, this OWE barrier is not solely determined by the final equilibrium temperature, but retains some dependence on the initial state, resulting in a complex, operator-dependent landscape on the Bloch sphere where the initial states are defined. In this context we are able to identify regions of anomalously low complexity, which are not related to the usual definition of "small quenches", and thus provide new scenarios in which current computational approaches can be used to predict the out-of-equilibrium dynamics. We thus present a further characterization of the possible scenarios for quantum advantage, by identifying some specific combinations of operators and initial states as scenarios that can be efficiently simulated classically. 

Moreover, we observe a distinct transition in the OWE as a function of the equilibration temperature of the initial state in a very narrow window. This behavior suggests a potential complexity transition in the equilibrium states, which we plan to investigate elsewhere.

The remainder of the manuscript is organized as follows: Section \ref{sec:II} introduces the notation and relevant quantities. In Section \ref{sec:III}, we decompose the operator into parallel and orthogonal components to the initial state, and analyze their complexity. Section \ref{sec:IV} charts the evolution of local operators on the Bloch sphere and defines the OWE; we plot the maximum OWE observed in our simulations as a function of the equilibration temperature and discuss its implications. Finally, we conclude with a summary and outlook in Section \ref{sec:V}.

\section{Setup and notation}\label{sec:II}
We study local quenches of the one dimensional Hamiltonian $\Ham$ defined as,
\begin{equation}
    \Ham=-\sum_j \big (\mathrm{J}~\sigma^z_{j}\sigma^z_{j + 1} + g~ \sigma^x_{j} + h~ \sigma^z_{j}\big ),
    \label{eq:Ising}
\end{equation}
For zero \textit{longitudinal field} $h=0$ and any \textit{transverse field} $g$ the system can be mapped through the Jordan-Wigner transformation to a chain of free Fermions, and it is thus \textit{integrable}; along this line, the system posses two phases: a disordered one for $\nicefrac{|g|}{J}>1$, and a ferromagnetic one elsewhere. The points $\nicefrac{|g|}{J}=1$ correspond to quantum critical lines within the universality class of the 2D Ising model. For the rest of values of the longitudinal field $h/J\neq 0$ the system is \textit{non-integrable} with a single disordered phase. Between the two critical lines, exactly at $\nicefrac{h}{J}=0$ the model displays a line of first order quantum phase transitions between two disordered phases in which the z-magnetization is predominantly aligned with the direction of the external field, but continuously vanishes as $|g|$ is increased. Throughout this work we will focus on the $g/J=1,~h/J=\nicefrac{1}{2}$ point, where the spectrum is known to be sufficiently chaotic; such a point in parameter space is expected to be particularly challenging for TN numerical simulations of the dynamics \cite{banuls2011strong, yang2020probing}.

We initialize the system as a \textit{Product State} (PS)
\begin{equation}
    \hat{\rho}(\theta,\varphi)=\bigotimes^L_{j= 1} \hat{\rho}_j(\theta, \varphi)= \bigotimes^L_{j= 1}|\theta,\varphi\rangle_{j}\langle \theta,\varphi |_{j},
    \label{eq:ps}
\end{equation}
where $|\theta,\varphi\rangle=\cos \frac{\theta}{2}\ket{0}+e^{i\varphi}\sin \frac{\theta}{2}\ket{1}$ is parameterised by the angles $\theta$ and $\varphi$ on the Bloch sphere.  These states can be thought as the ground states of single-spin translationally-invariant Hamiltonians $\hat{\mathcal{H}}_{\mathrm{1~spin}}=-f \sigma^{\parallel}_{\theta,\varphi} $ with strong fields $f\gg 1$ pointing towards a particular spatial direction $\Vec{n}_{\theta,\varphi}$ defining the linear combination of Pauli matrices $\sigma^{\parallel}_{\theta,\varphi}=\Vec{n}_{\theta,\varphi}\cdot\Vec{\sigma}$.

By virtue of the Eigenstate Thermalization Hypothesis \cite{srednicki1994chaos} we expect that at late enough times, the dynamics of local observables should converge to their expectation value in the thermal Gibbs ensemble, whose inverse temperature $\beta$ should be given implicitly by the solution of the following equation:  
\begin{equation}
\mathrm{Tr}\Bigg\{\hat{\mathcal{H}}\frac{e^{-\beta (\theta,\varphi) \hat{\mathcal{H}}}}{\mathrm{Tr}\{e^{-\beta (\theta,\varphi) \hat{\mathcal{H}}}\}}\Bigg\}=\mathrm{Tr}\{\hat{\mathcal{H}}\hat{\rho}(\theta,\varphi)\}.
    \label{eq:equilibration}
\end{equation}

\begin{figure}[t]
{\includegraphics[width=1.\columnwidth]{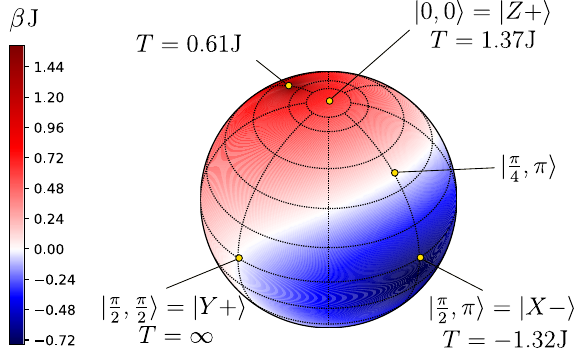}}
\caption{\textbf{Temperature map for PS defined on the Bloch sphere and the Ising Hamiltonian with $g=1$, $h=\nicefrac{1}{2}$}. The temperature values were obtained and benchmarked both from ED ($L\sim\mathcal{O}(10)$) and TN representations of the thermal density matrices ($L\sim\mathcal{O}(10^2)$). The diverging color scheme allows to identify the white band for which the temperature $T = \beta^{-1}$ becomes infinite and inverts its sign, separating a red patch with positive temperatures from the blue patch with population-inverted states and negative temperatures. The dynamics of representative states like $|X-\rangle$, $|Y+\rangle$ or $|Z+\rangle$ with negative, infinite and positive temperatures respectively has been tested in past works \cite{banuls2011strong, yang2020probing}, displaying a variety of thermalization regimes, some of them still on discussion. In our study we will also include other states like $|\frac{\pi}{4},~\pi\rangle$ with the purpose of rule out isolated behaviours due to symmetry.}
\label{fig:blochsphere}
\end{figure}

As a result from solving Eq. \ref{eq:equilibration} in the selected non-integrable point for angular steps $\Delta\theta, ~\Delta\varphi=1^\circ$, we generated a temperature map on the Bloch Sphere\footnote{We have used exact diagonalization for blocks of spins up to lengths $L=10$, and crossed-checked the stability of the results using MPO simulations up to $L=30$.}, which we present on Fig. \ref{fig:blochsphere}.

In this inverse temperature map of Fig. \ref{fig:blochsphere} we find an absolute maximum corresponding to the lowest temperature $\beta\mathrm{J}\simeq 1.630$ ($\nicefrac{T}{\mathrm{J}}=0.61$) close to the north pole $|0,~0\rangle =|Z+\rangle=|0\rangle$ and an absolute minimum with inverted population at $\beta\mathrm{J}\simeq -0.756$ ($\nicefrac{T}{\mathrm{J}}=-1.32$) around the point opposed to the external field, close to $|\frac{\pi}{2},~\pi\rangle =|X-\rangle=|-\rangle$. An infinite temperature $\nicefrac{T}{\mathrm{J}}=\infty$ isothermal line depicted in white contains $|\frac{\pi}{2},~\frac{\pi}{2}\rangle =|Y+\rangle=|\mathcal{R}\rangle$, and separates the region of positive temperatures (red coloring) from the region of negative temperatures (blue coloring). The dynamics of the aforementioned states has been extensively studied \cite{banuls2011strong, yang2020probing}, being $|0\rangle$ and $|\mathcal{R}\rangle$ examples of weak and strong thermalization, respectively: while \textit{weakly thermalizing} states only match their Reduced Density Matrices (RDMs) on small subsystems with the thermal RDM under time-averaging, \textit{strongly thermalizing} states bring their subsystem RDMs exponentially close to the exact thermal matrix. Other states, like $|-\rangle$, have triggered further discussion on whether they thermalize \cite{yang2020probing} or not \cite{banuls2011strong} in any of the former styles.

In the following we use TNs to compute the evolution of local observables in the Heisenberg picture; as a result we will be working with operators. A natural basis to express these operators is the \textit{Pauli string basis}, corresponding to all $4^L$ arrangements of Pauli matrices on $L$ sites $\sigma^{j_1}\otimes...\otimes\sigma^{j_L}$, with local operators at site $j$ drawn from the set $\{\mathds{1},~\sigma^x,~\sigma^y,~\sigma^z\}$. In such a basis an operator can be described as a \textit{Matrix Product Operator} (MPO) \cite{pirvu2010matrix} as 
\begin{equation} \sum_{j_1...j_L}\Tr{M^\alpha_{j_1}M^{\alpha\beta}_{j_2}...M^{\delta\gamma}_{j_{L-1}}M^{\gamma}_{j_L}}~\sigma_{j_1}...\sigma_{j_L},
\label{eq:MPO}
\end{equation}
\noindent where $M$ are rank $4\cdot \chi_{j,\mathrm{L}}\cdot\chi_{j,\mathrm{R}}$ tensors, and $\chi_{j,\mathrm{L/R}}$ are the bond dimensions to the left/right of the tensor sitting at site $j$.

\begin{figure}[t]
  \centering
  \includegraphics[width=\columnwidth]{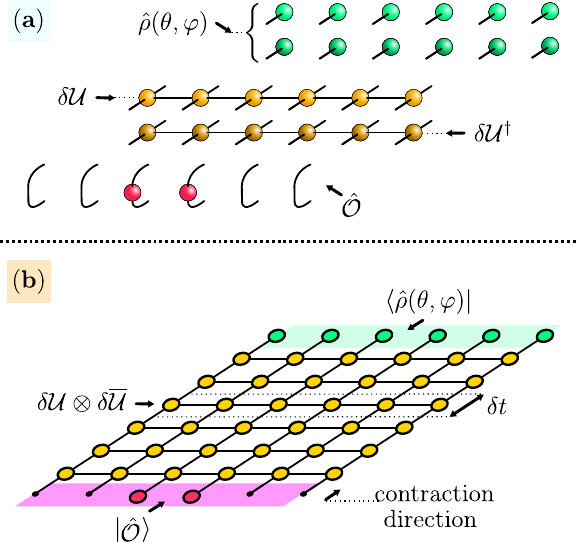}
  \caption{{\textbf{TN representation of an evolving local observable $\mathrm{Tr\{\hat{\rho}~\hat{\mathcal{O}}(t)\}}$.}} \textbf{(a) Elements of the quantum evolution.} We study the quench dynamics of initial translational invariant Product States described by the density matrix $\hat{\rho}(\theta,\varphi)$, parameterized by the Bloch sphere angles $\theta$ and $\varphi$; their vectorization is represented by a double layer of green tensors (light color for the \textit{ket}, and dark color for the \textit{bra}). These are evolved by repeated action of infinitesimal time translations on the \textit{ket} (forward evolution $\delta\mathcal{U}$, light yellow) and the \textit{bra} (backward direction $\delta\mathcal{U}^\dagger$, dark yellow). We consider only local observables, coupling the forward and backward layers with few non-trivial red tensors. \textbf{(b) Full TN to be contracted.} After flattening the forward and backward layers of the evolution, we obtain a 2D network with locally squared dimensions. Such a network can be contracted in several ways; here we choose the Heisenberg picture evolution in which rows of folded evolution MPOs are absorbed into the operator edge.}
  \label{fig:evolution}
\end{figure}

The main equation we will encode with TNs is
\begin{equation}
\expval{\Op(t)} = 
 \Tr{\Rhop (0) ~e^{+i\hat{\mathcal{H}}t}~\Op ~e^{-i\hat{\mathcal{H}}t}},
 \label{eq:evolution}
\end{equation}
where $\Rhop (0)=\Rhop (\theta,\varphi)$ is the density matrix representing the intial state.
In order to use the standard TNs techniques we have to divide the evolution into small time steps $\delta t$, which together with the locality of $\Ham$ allows to approximate the full evolution operator as a sequence of infinitesimal\footnote{For each step $\delta t$ we use a second-order Trotter-Suzuki approximation to $\delta\hat{\mathcal{U}}=\hat{\mathcal{U}}(\delta t)$ \cite{suzuki1990fractal, suzuki1992fractal}, which introduces an accumulated error $\mathcal{O}(t\cdot\delta t^2)$. For other optimized versions of the evolution operator, we refer the reader to \cite{zaletel2015timeevolving, barthel2020optimized}.} rotations $\delta \hat{\mathcal{U}}=e^{-i\Ham \delta t}$. Crucially each $\delta \hat{\mathcal{U}}$ can be cast as an MPO with finite bond dimension and, as a result, all the objects entering the evolution in Eq. \eqref{eq:evolution} can be encoded by MPOs: the initial density matrix $\Rhop (0)$, the time step translation $\delta\hat{\mathcal{U}}$ and the evolving operator $\Op$, as illustrated in Fig. \ref{fig:evolution} (a); together they form the 2D TN to be contracted. The complexity of the evolution is thus translated to that of contracting such a TN.

Given the trace in Eq. \ref{eq:evolution} the TN representation becomes a double layer of tensors, one for the \textit{ket} (the forward evolution $\hat{\mathcal{U}}$) and another for the \textit{bra} (the backward evolution $\hat{\mathcal{U}}^\dagger$), closed by the operator edge\footnote{Note that in the case of studying initial mixed states, the forward and backward layers would be coupled at the edge corresponding to the state, too.}, as depicted in Fig. \ref{fig:evolution} (a). We flatten the double layer as in Fig. \ref{fig:evolution} (b), where we group the upper and lower tensors of Fig. \ref{fig:evolution} (a) together and merge the forward and backward legs of the tensors.

In this setting, operators and density matrices become vectors that we encode as \textit{Matrix Product States} (MPS) 
\begin{equation}
\sum_{j_1...j_L}\Tr{M^\alpha_{j_1}...M^{\gamma}_{j_L}}~|\sigma_{j_1}...\sigma_{j_L}\rangle,
     \label{eq:MPS}
\end{equation}
and are  placed at the edges of the 2D flat TN depicted in Fig. \ref{fig:evolution} (b). 

Working in the Heisenberg picture involves to start contracting the TN from the operator and proceeding towards the initial state, which will be the last piece to be contracted. Given the flattening, each step is performed by applying the folded version of the unitary evolution \cite{muller2012tensor}, which recasts the super-operator $\hat{\mathcal{U}}(t)~(~\cdot~)~\hat{\mathcal{U}}^\dagger(t)$ as a single-layer object $\hat{\mathcal{U}}(t)\otimes\hat{\overline{\mathcal{U}}}(t)$ and thus as an MPO applied on the vectorized operator,
\begin{equation}
    \ket{\Op (t)} = \hat{\mathcal{U}}(t)\otimes\hat{\overline{\mathcal{U}}}(t) ~ \ket{\Op (0)}.
    \label{eq:Heisen_pict}
\end{equation}

Since at each step the evolving operator is encoded as a MPS, we measure the computational complexity of simulating the Heisenberg evolution using the \textit{Operator Space Entanglement Entropy} (OSEE) \cite{prosen2007efficiency, prosen2007operator, znidaric2008manybody, pizorn2009operator, dubail2017entanglement, zhou2017operator}, which is the entanglement entropy of the vector encoding the evolved operator. Considering the iterative loss of operator norm due to truncation, the OSEE is computed through the retained singular values $\lambda_{j,a}$ at each bipartition $j$:
\begin{equation}
\begin{split}
    \mathrm{OSEE}(j,~ t)=-\sum_a\frac{\lambda_{j,a}^2(t)}{1-\varepsilon (t)}\log \bigg [ \frac{\lambda_{j,a}^2(t)}{1-\varepsilon (t)} \bigg ],
\end{split}
\label{eq:OSEE}
\end{equation}
 where \begin{equation}
\varepsilon (t)=\sum_j\sum^{\chi^*}_{a=1}\lambda^2_{j,a}=1-||\Op (t)||_2,
\label{eq:norm_err}
\end{equation} and the norm is 
\begin{equation}
    ||\Op||_2=\braket{\Op}{\Op} = \frac{1}{2^L}\mathrm{Tr}\{\hat{\mathcal{O}}^{\dagger}\hat{\mathcal{O}}\}.
    \label{eq:innerprod}
\end{equation}

In practice, at each time step one MPO encoding the vectorized evolution $\delta \hat{\mathcal{U}}\otimes \delta \hat{\overline{\mathcal{U}}}$ is absorbed into the boundary state obtained at the previous step, evolving from $t$ to $t+\delta t$. The resulting MPS bond dimensions increase, and we use a TEBD \cite{vidal2003efficient, vidal2004efficient} truncation scheme to keep it under control. This strategy is doomed to fail after relatively short times, since the bond dimension of the state encoding the operator increases exponentially with the time for a fixed accuracy, as a consequence of the linear growth of the OSEE for generic scenarios \cite{prosen2007operator}. This means that all the results we present are obtained for relatively short times, since the maximum time within reach of our simulations only increases logarithmically with the amount of computational resources we dedicate to them.

\section{Parallel basis induced by an initial product state}\label{sec:III}

In the Heisenberg picture the operator is evolved independently of the initial state. When computing expectation values however we just focus on specific subsets of matrix elements of the full operator superposition. It is thus natural to address the growth of complexity of these specific matrix elements. This is what we do here by introducing the notion of parallel Pauli strings. 
Since all matrix elements depend on time, different elements to the ones selected by the initial state at a given time are  later rotated to the parallel subspaces during the dynamics. It is thus also important to understand such processes, and we address them here in the section about backflow.

As mentioned in Sec. \ref{sec:II}, each PS has a different equilibration temperature, so here we will focus on a pair of states which illustrates the two limiting case studies: the first one, $|0,~0\rangle$, corresponds to a low temperature $\nicefrac{T}{\mathrm{J}}=1.37$, and the second one to a higher temperature, $|\frac{\pi}{4},~\pi\rangle$ ($\nicefrac{T}{\mathrm{J}}\gg 1$). In this section our analysis suggests that in the Heisenberg picture states with corresponding high equilibration temperatures appear to contain less Pauli weight along their dynamics, a feature that at the same time is shown to be connected to a lower computational complexity; we defer a full analysis of the temperature dependence of such complexity to the later section \ref{sec:IV}.

We start by rewriting the initial PS in the Pauli string basis. Following the same vectorization scheme as for operators in Fig. \ref{fig:evolution} and Eq. \ref{eq:MPS}, the PS from Eq. \ref{eq:ps} can be written as a superposition of strings
\begin{equation}
\begin{split} 
    &|\hat{\rho}(\theta,\varphi)\rangle=\frac{1}{2^L}\bigotimes^L_{j= 1} \bigg ( |\mathds{1}_j\rangle + |\sigma^{\parallel}_j\rangle \bigg )\quad\mathrm{with}
    \\
    &|\sigma^{\parallel}_j\rangle = 
    \sin \theta (\cos\varphi|\sigma^x_j\rangle +\sin\varphi|\sigma^y_j\rangle) +\cos\theta|\sigma^z_j\rangle.
\end{split}
   \label{eq:sigma_parallel}
\end{equation}
\noindent where we define the \textit{parallel matrices} $\sigma^{\parallel}$ to a given set of angles defining the initial state in the Bloch sphere, and drop the angular subindices from here on. A complete, state-specific basis $\{\mathds{1},~\sigma^{\parallel},~\sigma^{\perp,1}, ~\sigma^{\perp,2}\}$ \footnote{Note that for an initial inhomogeneous PS we could use a spatially dependent parallel basis $\{\mathds{1},~\sigma^{\parallel}_j,~\sigma^{\perp,1}_j, ~\sigma^{\perp,2}_j\}$.} is found through orthogonality:
\begin{equation}
     \langle \sigma^{\parallel} | \sigma^{\perp,i}\rangle = 0\quad \mathrm{for}\quad i=1,2.
    \label{eq:string_superpos_ortho}
\end{equation}

Given the definition of the \textit{Pauli weight} $\omega$ using Eqs. \ref{eq:sigma_parallel} and \ref{eq:string_superpos_ortho}, we split it as the sum $\omega=\omega^{\parallel}+\omega^{\perp}$ of \textit{parallel weight} $\omega^{\parallel}$ and \textit{orthogonal weight} $\omega^{\perp}=\omega^{\perp, 1}+\omega^{\perp,2}$ whenever we take an initial PS as a reference. 

For example, given the initial state $|0\rangle$, the parallel weight is given by $\sigma^z$, while the orthogonal weight is hosted by $\sigma^{x,y}$. Despite the total weight $\omega$ of a given string is preserved by local basis rotations mixing Pauli matrices, the relative weights $\{\omega^{\parallel},~\omega^{\perp, 1},~\omega^{\perp, 2}\}$ do change. For PS, the \textit{parallel basis} is the one for which all the Pauli weight is of parallel character: $\omega = \omega^{\parallel}$. Such a basis is encoded in the local rotation $\mathcal{B}_{\theta,\varphi}$ that diagonalizes the initial state.

As a result we can  split the operator into two parts: in first place a \textit{contributing superposition} $|\hat{\mathcal{O}}^{\mathrm{c}}(t)\rangle$ containing only parallel weight ($\omega^{\perp}=0$) that yields the non-zero part of the overlap, and in second place a \textit{non-contributing superposition} $|\hat{\mathcal{O}}^{\mathrm{nc}}(t)\rangle$ with any parallel content ($\omega^{\parallel}>0$) but at least one orthogonal insertion ($\omega^{\perp}\geq 1$), that cancels out in the expectation value
\begin{equation}
\begin{split}
    &|\hat{\mathcal{O}}(t)\rangle = |\hat{\mathcal{O}}^{\mathrm{c}}(t)\rangle + |\hat{\mathcal{O}}^{\mathrm{nc}}(t)\rangle,\\
    &\langle \hat{\rho} (\theta,\varphi)|\hat{\mathcal{O}}^{\mathrm{nc}}(t)\rangle = 0.
\end{split}
    \label{eq:op_contr_nocontr}
\end{equation}

Via standard MPO techniques that are detailed in the Appendix \ref{app:MPOs}, we can furthermore characterize the role of the parallel and orthogonal superpositions by resolving their Pauli weight. To start such analysis, we study the splitting the total operator norm  $\langle \hat{\mathcal{O}}(t)|\hat{\mathcal{O}}(t)\rangle$ into a \textit{contributing} $\rho^{\mathrm{c}}(t)$ and a \textit{non-contributing} $\rho^{\mathrm{nc}}(t)$ \textit{densities}
\begin{equation}
\begin{split}
    \rho^{\mathrm{c}}(t) &= \langle \hat{\mathcal{O}}(t) | \hat{\mathcal{O}}^{\mathrm{c}}(t) \rangle, \\
    \rho^{\mathrm{nc}}(t) &= \langle \hat{\mathcal{O}}(t) | \hat{\mathcal{O}}^{\mathrm{nc}}(t) \rangle, \\
    \rho^{\mathrm{c}}(t) + \rho^{\mathrm{nc}}(t) &= 1-\varepsilon(t).
    \label{eq:wcdens}
\end{split}
\end{equation}

\begin{figure}[!htb]
  \includegraphics[width=1.\columnwidth]{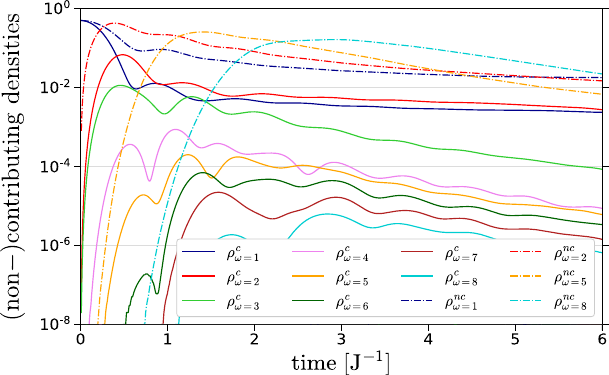}
  \caption{\textbf{Evolution of the (non-)contributing densities for the initial bulk magnetization $\sigma^x$}. For the initial state $|\frac{\pi}{4},\pi\rangle$, we observe the general separation between the leading non-contributing norm $\rho^{\mathrm{nc}}$ (dash-dotted curves) and the sub-leading contributing one $\rho^{\mathrm{c}}$ (solid curves). The orthogonal subspace ($\omega^{\perp}\geq 1$) keeps transferring norm to longer strings, as observed from the sequential peaks and decays for consecutive total weights. Opposed to that, the parallel densities ($\omega^{\perp}=0)$ progressively decay, hosting a small fraction of the total norm. As expected for an initially local operator, the lightest weights initially dominate the norm, but soon strings of weight $\mathcal{O}(10)$ proliferate both in the parallel and orthogonal superpositions around the same times. In this specific scenario, the weight of the heavy strings in the contributing subspace is however strongly suppressed.}
 	\label{fig:globalrhos}
\end{figure}

In Fig. \ref{fig:globalrhos} we plot the early time behavior of the Pauli-weight-resolved (non-)contributing densities for the evolved bulk local x-magnetization\footnote{Our exhaustive simulations have not found qualitative differences with $\sigma^{y,z}$.} $\sigma^x$. We depict the results for the initial state $|\frac{\pi}{4},\pi\rangle$ in a lattice of size $L=30$, evolved with TEBD for $\delta t \mathrm{J} = 0.01$, tail truncation $\lambda^2 < 10^{-10}$ and maximum bond dimension $\chi =576$. Time is in the horizontal axis, while the superposition weight of each subspace labelled by Pauli weight $\omega$ is plotted on the vertical axis. The solid lines represent the evolution of the contributing densities $\rho^{\mathrm{c}}_\omega$, while the dash-dotted lines stand for the norms of non-contributing subspaces $\rho^{\mathrm{nc}}_\omega$.

We first note that already for times of $\mathcal{O}(1)$ in units of the inverse interaction $\mathrm{J}^{-1}$ the superposition harvests heavy strings with Pauli weight $\mathcal{O}(10)$ (see dark red and cyan curves with $\omega=7,~8$ respectively).

In second place, we detect a difference in the order of magnitude between contributing and non-contributing densities: the discontinuous lines fall on the window between 0.01 and 1 for the simulated times, while the solid lines decay directly to 0.01 and below; this means that only a tiny piece of the total operator norm is needed for the computation of the expectation value at a given time. 

The third observation is that strings within the contributing subspace tend to host less and less norm, both as $\omega$ and $t$ grow (compare the curves $\rho^{\mathrm{c}}_{\omega=1, 2, 5, 8}$). This result is coherent with the simulation setup: the initial operator is a single string with Pauli weight $\omega=1$, and the Hamiltonian is a nearest-neighbour model that couples non-trivially only the closest sites, forcing the appearance of heavier strings to be sequential with $\omega$ and $t$; on top of that, the unitary evolution will keep scrambling the norm until the maximal weight $\omega=L$ set by the finite system simulation is reached, explaining the continued overall decay of the densities after transitory oscillations. 

The contributing superposition appears to display different relaxation rates for the lightest strings $\omega \leq 2$ (fast-relaxing) and heavier ones $\omega\geq 3$ (slow-relaxing). This separation of time scales appears to be related to the conservation of energy, as pointed out in \cite{khemani2018operator}, and to the fact that the initial operator has its Pauli weight localized in few sites.

Finally, the direct decay of the contributing densities contrasts with the sequential population of heavier orthogonal subspaces: the superposition weights of the latter peak at a given time for a given Pauli weight $\omega$ and decay afterwards, transmitting most of its norm to the following subspace with weight $\omega+1$ as the time advances; this can be appreciated for $\rho^{\mathrm{nc}}_{\omega=1, 2, 5, 8}$.

        \subsubsection{Direct contributions from parallel subspaces}\label{ssec:direct}

So far we found that the contributing norm soon spreads to heavy Pauli strings which  host a reduced fraction of the norm. The final expectation value  requires summing these contributions, which include relative signs and as such knowing just their weights in the operator norm does not allow to deduce their relative importance. On top of it, heavy strings with fixed weight can be arranged in exponentially many spatial distributions given by combinatorics.

 In order to properly visualize the contributions to the local  expectation value, we will resolve the contributions $\mathcal{O}_{\omega}(t)$ by their Pauli weight:
\begin{equation}
\begin{split}
    \mathcal{O}(t)&=\langle \hat{\rho} (\theta,\varphi)|\hat{\mathcal{O}}(t)\rangle =\\&= \sum_{\omega} \langle \hat{\rho} (\theta,\varphi)| \hat{\mathcal{P}}_{\omega} |\hat{\mathcal{O}}^{\mathrm{c}}(t)\rangle=\sum_{\omega}\mathcal{O}_{\omega}(t).
\end{split}
    \label{eq:contribs}
\end{equation}
were the projector $\hat{\mathcal{P}}_{\omega}$ onto strings with Pauli weight $\omega$ and can be efficiently represented as an MPO and (see Appendix \ref{app:MPOs}).

In Fig. \ref{fig:direct} we present the absolute values $|\mathcal{O}_{\omega}(t)|$ for the operator $\sigma^x_{j=\frac{L}{2}} (t)$ and the initial states: $|0,~0\rangle$ and $|\frac{\pi}{4},~\pi\rangle$. The simulations were run with $\delta t \mathrm{J} = 0.01$, and at each step we truncate the operator entanglement spectrum discarding those eigenvalues fulfilling $\lambda^2 < 10^{-10}$. Furthermore we set a maximum bond dimension $\chi = 256$ that soon gets saturated. 

\begin{figure}[!htb]
\includegraphics[width=1.\columnwidth]{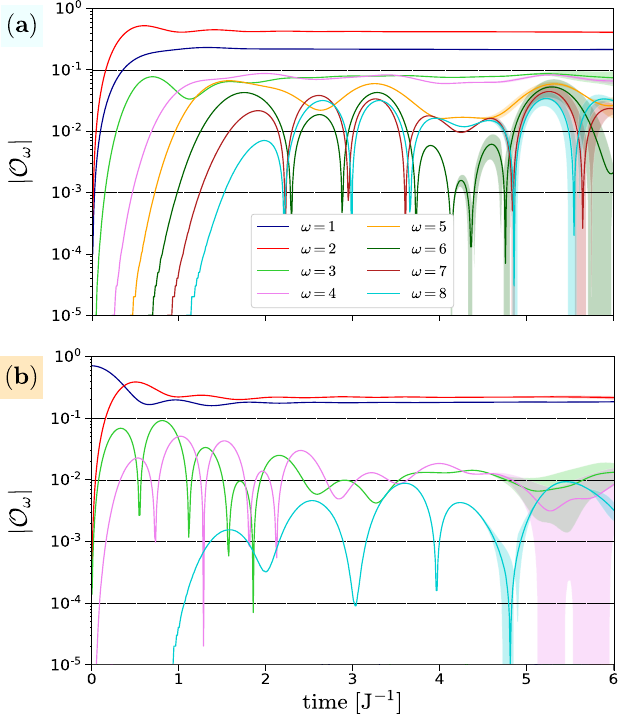}
  \caption{\textbf{Contributions to local observables from fully parallel strings for states $|0,~0\rangle$ (top panel) and $|\frac{\pi}{4},~\pi\rangle$ (bottom panel)}. While all the considered strings yield relevant contributions for the initial state $|0,~0\rangle$ (until $\omega=8)$ in panel \textbf{(a)}, a gap of an order of magnitude opens between $\omega\leq 2$ and $\omega\geq 3$ for the state $|\frac{\pi}{4},~\pi\rangle$ in panel \textbf{(b)}; generically, the size of the contributions decreases with $\omega$, and heavier spaces tend to delay the stabilization of their contributions, as expected from local initial operators. The shaded areas represent the difference between $\chi=256$ and $\chi'=\frac{\chi}{2}$, making evident that heavier subspaces suffer the most from truncation.}
 	\label{fig:direct}
\end{figure}

As a general theme, we observe a decay of $|\mathcal{O}_{\omega}(t)|$ as $\omega$ increases, in accordance with their corresponding densities in Fig. \ref{fig:globalrhos}, a footprint the locality of the initial operator. 

We also observe a sequential stabilization of the contributions from each subspace: $\mathcal{O}_{\omega=1, 2}$ yield an almost stationary signal at some early time $t^{\mathrm{sat}}_{1, 2}$, while $\mathcal{O}_{\omega=3, 4}$ take longer than our simulation times to stabilize; we expect a similar trend for increasing $\omega$. On top of that, subspaces appear to saturate in pairs (e.g. $\omega=1,~2$ and $\omega=3,~4$), possibly a consequence of the underlying conservation of the powers\footnote{$\Ham$ contains strings with at most weight 2, $\hat{\mathcal{H}}^2$ contains until $\omega=4$, etc.} of the local Hamiltonian \cite{khemani2018operator}. 

Since our results are obtained with simulations that encode the  operator into a MPS with finite bond dimension, it is necessary to establish their regime of validity. This requires understanding which subspaces are most vulnerable to the compression from the TEBD method. With this purpose, we introduce a shaded error region indicating the difference between the values with $\chi=256$ and with $\chi=128$. By observing the size of these shaded regions in Fig. \ref{fig:direct} , we conclude that the contributions of light strings ($\omega < 5$) are pretty stable, while those from heavier strings ($\omega\geq 5$) are more sensitive to the truncation in both states: see for example the shaded areas in the upper panel for weights $\omega=6$, 7 and 8, depicted in dark green, dark red and cyan respectively. This means that heavy string contributions will receive stronger corrections upon increasing the maximum $\chi$; on the level of the OSEE, this implies that heavy parallel operators are relegated to the tail of the distribution of the operator Schmidt values, and as such they are the first ones discarded by the TEBD algorithm. The latter can be restated as an overall inverse correlation between the support of a string on the leading operator Schmidt vectors and the parallel Pauli weight $\omega^{\parallel}$ of that string.

This is a useful piece of information in combination with the hierarchical reduction of the contributions with increasing $\omega$: heavy subspaces accumulate most error in TEBD simulations, but at the same time contribute less and less to the observable. This suggests that the generic TEBD should be able to predict accurately the dynamics given some threshold of retained Pauli weight $\omega^*$. The key point is to understand how large $\omega^*$ needs to be in order to guarantee accurate predictions. This brings us to the difference between the upper panel and the lower one in Fig. \ref{fig:direct}: for the state $|0,~0\rangle$ the contributions oscillate wildly across orders of magnitude, and heavy strings with weight $\omega=8$ (cyan)  contribute similarly to strings with weight $\omega=5$ (yellow) or $\omega=3$ (green). On the other hand, in the lower panel an exponential gap appears between the lightest strings $\omega \leq 2$ and the heavier ones.  In the lower panel one could thus truncate the heavier strings, while in the upper one needs to retain all the strings we have analyzed due to the lack of a clear hierarchy among them. The accurate simulation of the x-magnetization is thus cheaper for the lower panel, where the heavy subspace only contributes as a small correction to the leading light string contributions. The simulation in the upper panel could not be simplified in this regime, resulting in generically higher bond dimensions and numerical cost. It is thus computationally harder to produce results with the same level of accuracy for the quench of the upper panel than for that in the lower panel.

These results seem to suggest a connection between the equilibrating temperature of the dynamics and the Pauli weight of the relevant contributions. $|0,~0\rangle$ corresponds to a low temperature $\nicefrac{T}{\mathrm{J}}=1.37$, and clearly requires heavier strings to represent its dynamics, whilst the higher temperature state $|\frac{\pi}{4},~\pi\rangle$ ($\nicefrac{T}{\mathrm{J}}\gg 1$), needs only short strings to be properly simulated. In Sec. \ref{sec:IV} we will present an extensive analysis to verify if such a connection between Pauli weight and equilibration temperature exists, leading us to unveil a very complex scenario.

        \subsubsection{Backflow from the orthogonal subspaces}\label{ssec:backflow}

Before addressing the possible connection between equilibration temperature and complexity of the evolution, in this section we analyze the role of the orthogonal components of the operators in the dynamics. As explained in the former Sec. \ref{sec:III}, orthogonal components have zero overlap with the initial PS defining the parallel basis. Nevertheless, the norm of these strings varies with time, meaning that they can be rotated back by the evolution into the contributing subspaces, a process that we dub as \textit{backflow} of orthogonal norm into the parallel subspaces. 

To determine the magnitude of this backflow, we design the following protocol: in first place, we evolve an initially local operator and monitor the norm deposited onto strings with fixed orthogonal weight $\omega^{\perp}$. The norm of this orthogonal superposition reaches a peak at some weight-dependent time $t_0$ and later decays, transmitting its norm to the following weight $\omega^{\perp}+1$. At the time of the peak $t_0$, we project out the rest of the superposition with $\hat{\mathcal{P}}_{\omega^{\perp}}=\hat{\mathcal{P}^{\mathrm{nc}}}\hat{\mathcal{P}}_{\omega=\omega^{\perp}}$ (we refer to the Appendix \ref{app:MPOs} for the detailed MPO implementation), and further evolve this non-contributing vector from $t_0$ while monitoring the overlap with the state. Such an overlap gives us an estimation of the return of norm from irrelevant orthogonal spaces to contributing ones:
\begin{equation}
\begin{split}
    \mathcal{O}^{\mathrm{back}}_{\omega^{\perp}}(t)=\langle \rho(\theta, \varphi) |[\mathcal{U}\otimes\overline{\mathcal{U}}](t-t_0)\hat{\mathcal{P}}_{\omega^{\perp}}|\hat{\mathcal{O}}(t_0)\rangle.
    \label{eq:protocol}
\end{split}
\end{equation}

\begin{figure}[!htb]
\includegraphics[width=1.\columnwidth]{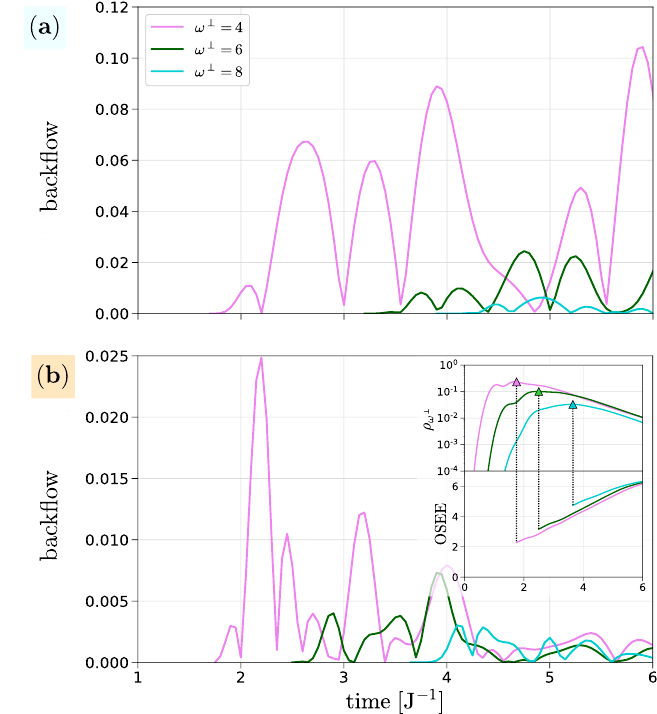}
  \caption{\textbf{Backflow from orthogonal subspaces for initial states $|0,~0\rangle$ (upper row) and $|\frac{\pi}{4},~\pi\rangle$ (lower row)}. Estimation of the backflow from orthogonal strings following the protocol in Eq. \ref{eq:protocol}: starting from the peak norm for strings with fixed $\omega^{\perp}=4,~6,~8$ (upper-right insets), we evolve the initially orthogonal superposition at time $t_0$ to see the return to the contributing space with $\omega^{\perp}=0$; the corrections tend to decrease with $\omega^{\perp}$. The upper state in panel \textbf{(a)} gets important corrections of order $\mathcal{O}(10^{-1})$ from the smallest orthogonal weight $\omega^{\perp}=4$ (solid violet curves), whereas the lower state in panel \textbf{(b)} only receives $\mathcal{O}(10^{-2})$, which indicates that the backflow is more prominent on $|0,~0\rangle$ than in $|\frac{\pi}{4},~\pi\rangle$; for the biggest simulated weight ($\omega^{\perp}=8$), the backflow is completely negligible as compared to other sources of error. The validity of the results is restricted to the regime of linear growth of the OSEE from each superposition (lower-right inset).}
 	\label{fig:backflow}
\end{figure}

In Fig. \ref{fig:backflow} we show the results of simulating $|\mathcal{O}^{\mathrm{back}}_{\omega^{\perp}}(t)|$ for $\omega^{\perp}=4,$ 6, and 8 for the same initial states $|0,~0\rangle$ and $|\frac{\pi}{4},~\pi\rangle$ as in Sec. \ref{ssec:direct}, with numerical parameters $\delta t\mathrm{J} = 0.05$, $\lambda^2<10^{-10}$ and $\chi=768$. Generically, we detect a suppression of backflow as $\omega^{\perp}$ increases, meaning that non-contributing strings with orthogonal weight above a threshold $\omega^*$ may be suppressed with a reduced impact on the dynamics. 

The inset in the lower panel refers to the protocol design: the densitites with a given orthogonal content are monitored until they peak at some time $t_0$, and the retained orthogonal superpositions are left to evolve. In the lower inset the OSEE of the orthogonal superpositions is depicted, clearly growing linearly with time; the apparent saturation of this measure is due to the limited bond dimension used on the simulation, limiting the times of faithful simulation.

While in the upper panel we observe a strong backflow from $\omega^{\perp} = 4$ (violet line) contributing to the observable on the order $\mathcal{O}(10^{-1})$, for heavier strings like $\omega^\perp=8$ (cyan line) this effect is suppressed. Interestingly enough, the behaviour of the high temperature state in the lower panel with respect to the low temperature one in the upper panel reproduces what we observed already in Fig. \ref{fig:direct}: the backflow from $\omega^{\perp} = 4$ in state $|\frac{\pi}{4},~\pi\rangle$ is already an order of magnitude below the same backflow for $|0,~0\rangle$. Such a result points towards the fact that some states will require keeping heavier Pauli strings in the evolving Heisenberg operators, like $|0,~0\rangle$, while the evolution of local observables for others may be faithfully reproduced with TEBD or simple modifications around it, like $|\frac{\pi}{4},~\pi\rangle$. In addition, this trend appears to be connected to the equilibration temperature of the initial state in the Gibbs ensemble.

Importantly enough, the lower state shows also a decay in time of the backflow; despite the simulated times do not allow to visualize this in the upper panel, we expect the same behavior on longer timescales, since we know that the orthogonal strings lose norm to heavier spaces (see Fig. \ref{fig:globalrhos}), which will eventually backflow with lower intensity. These results clearly suggest that the suppression of heavy orthogonal subspaces during the TEBD evolution would have an impact below the accuracy marked by both, truncation and Trotter errors. Thus we expect that the aforementioned threshold $\omega^*$ needed to simulate the local dynamics of some state will be fixed by the size of the direct contributions in terms of $\omega$, and not by the backflow to the diagonal basis.

\section{Charting the operator complexity of the Heisenberg evolution on the Bloch sphere}\label{sec:IV}

In light of the results of the previous section we can safely ignore the backflow of the orthogonal subspace and focus on the buildup of Pauli weight in the contributing subspaces. We can now systematically characterize how the required Pauli weight of the direct contributions depends on the initial states, and how this connects to the hardness of the simulation. With this purpose we introduce the concept of Operator Weight Entropy, that will help us charting the complexity of the evolution for an extensive set of initial states on the Bloch sphere. In doing so, we unveil an abrupt change in complexity of simulating the local dynamics, which occurs within the same window of temperatures for a variety of sets of initial states.

    \subsection{Operator Weight Entropy}\label{ssec:OLEE}

So far we have seen that the Pauli-weight-resolved contributions to the expectation values of local operators strongly depend on the initial state. On top of it, contributions from heavy strings are less stable against truncation, implying that states with stronger presence of heavy strings will be more difficult to simulate with TEBD. 

In order to perform a systematic comparison of the complexity of the specific matrix elements of the evolved operator selected by the initial states, we would like to define a single quantifier of such complexity that informs us on the Pauli string content of the operator sorted by the Pauli weight. 

We start by computing the truncated expectation value of the operator $\mathcal{O}(t)$ over the desired initial state, including only strings with maximum Pauli weight $\omega$, i.e. the \textit{accumulated observable sum} $\mathcal{O}^{\mathrm{acc}}_{\omega}(t)$ defined by limiting the weight included in the sum of Eq. \eqref{eq:contribs} up to a maximum $\omega$, 
\begin{equation}
    \mathcal{O}^{\mathrm{acc}}_{\omega}(t)=\sum_{n=0}^{\omega}\mathcal{O}_{n}(t).
\end{equation}
The exact observable is written as $\mathcal{O}(t) = \mathcal{O}^{\mathrm{acc}}_{\omega=\infty}(t)$, and the accumulated observable allows us to define the instantaneous deviation between the exact and the $\omega$-truncated expectation value:
\begin{equation}
    d_{\omega} (t) = |\mathcal{O}(t) - \mathcal{O}^{\mathrm{acc}}_{\omega}(t)|.
\end{equation}

Once we have computed these distances we can use them to build a probability distribution by normalizing each of the distances using $ \mathcal{N}_{\omega*}(t)$ their sum up to a cut-off $\omega^*$, fixed by our computational resources\footnote{Recall that the bond dimension of the MPO projector onto strings with weight $\omega$ grows linearly with it (see App. \ref{app:MPOs}) and thus we cannot take $\omega^*$ excessively large without incurring in a prohibitive computational cost.}
\begin{equation}
    \mathcal{N}_{\omega*}(t) = \sum_{\omega=0}^{\omega*} d_{\omega}(t).
\end{equation}

The \textit{normalized deviation instantaneous distribution} reads
\begin{equation}
    p_{\omega,\omega^*}(t)=\frac{d_{\omega, t}}{\mathcal{N}_{\omega*}(t)}.
\end{equation}

The distribution $p_{\omega,\omega^*}(t)$ gives us information about the relevance of strings with Pauli weight smaller or equal to $\omega$ in the scenario in which the operator is truncated to string superpositions with at most $\omega^*$. We characterize this through the entropy of this distribution, which we define as \textit{Operator Weight Entropy} (OWE):
\begin{equation}
    \mathrm{OWE}(t)=- \sum_{\omega=0}^{\omega*} p_{\omega,\omega^*}(t) \log_2 p_{\omega,\omega^*}(t).
\end{equation}

Given the cumulative definition of $p_{\omega,\omega^*}$, we need to carefully consider the possible behaviours of the OWE across time. For example, an observable that is accurately represented by a finite $\omega^*$ for arbitrarily late times would have an OWE that saturates after an initial growth. Opposed to that, an operator whose contribution is equally distributed on all possible accessible strings given the locality of the dynamics, would have an OWE that grows logarithmically in time\footnote{$p_{\omega,\omega^*}$ would be linearly decreasing for all accessible $\omega$ given its cumulative definition.}. A linearly increasing OWE would only be obtained by an operator that is described by many Pauli strings whose weight exceed $\omega^*$, since only in that case all the $p_{\omega,\omega^*}$ would be constant.

\begin{figure}[!htb]
  \includegraphics[width=1.\columnwidth]{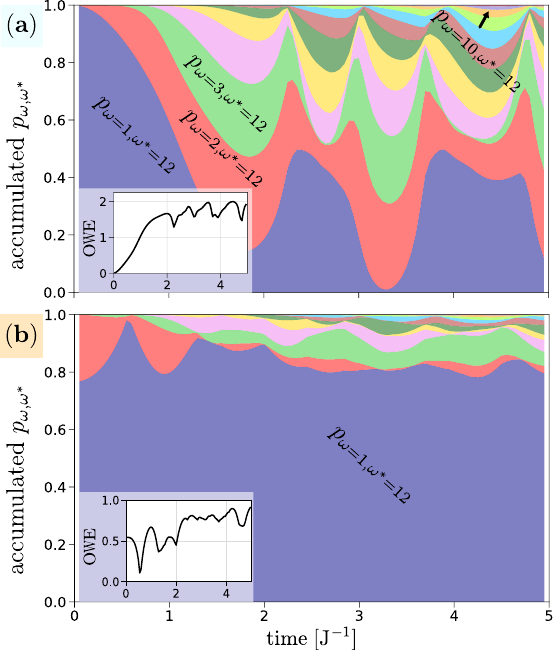}
  \caption{\textbf{Evolution of the probability distribution $p_{\omega,\omega^*=12}(t)$ induced by the distance between the exact x-magnetization and the truncated Pauli weight sum for initial states $|0,~0\rangle$ and $|\frac{9\pi}{10},~0\rangle$}. The probabilities are laid one on top of each other, such that a vertical cut at a fixed time unveils the percentages due to truncating the Pauli weight above the respective $\omega$. We start from a local operator that has weight 1, as witnessed by the fully  blue band. {\bf{(a)}} For the initial state $|0,~0\rangle$ the dynamics soon harvests contributing strings with weights 2 (red band), 3 (green band) and up to $\omega = 10$ (narrow orange band) for an accurate description. {\bf{(b)}} Other states like $|\frac{9\pi}{10},~0\rangle$ converge in the Pauli weight sum much faster: in this case, strings of weight 1 (blue band) dominate the observable for all the simulated times. As an inset in each panel we present the OWE of each distribution, which stabilizes at an almost doubled value for {\bf{(a)}} than for {\bf{(b)}}, reflecting the stronger requirement of Pauli weight in the first case.}
 	\label{fig:histograms}
\end{figure}

In Fig. \ref{fig:histograms} we present the time evolution of the distribution $p_{\omega,\omega^*}(t)$ in cumulative form: consecutive weights appear one on top of another, such that a vertical cut of the plot contains the instantaneous values of the distribution such that $\sum_{\omega = 1}^{\omega^*}p_{\omega,\omega^*}=1$. We choose as initial states $|0,~0\rangle$ (upper panel, with equilibration temperature $\nicefrac{T}{\mathrm{J}}=1.37$) and $|\frac{9\pi}{10},~0\rangle$ (lower panel, with equilibration temperature $\nicefrac{T}{\mathrm{J}}=3.25$) and bulk x-magnetization, since they illustrate extreme cases from the exhaustive study presented in Fig. \ref{fig:40states}. The numerical parameters are $\delta t \mathrm{J} = 0.05$, $\lambda^2<10^{-10}$ and $\chi=576$. The observable defined  by a single $\omega=1$ term (blue filling), and the heavier strings start contributing at early stages most noticeably for the state $|0,~0\rangle$ in the upper panel, which reaches non-negligible contributions from weight $\omega\sim\mathcal{O}(10)$ by $t\mathrm{J}=4$. On the other hand, the state $|\frac{9\pi}{10},~0\rangle$ presented in the lower panel  picks up almost immediately its small $\omega=2$ contributions, but nevertheless stays dominated by light strings.

The clear difference between the two panels in Fig. \ref{fig:histograms} signals once more the existence of states which are much easier to simulate, and here again this points to the fact that such states are those in which lighter Pauli strings dominate. We remind the reader that in Fig. \ref{fig:direct} we have shown that heavier strings have support on the tail of the Schmidt vectors of the operator, and thus their evolution is strongly affected by the truncation in TEBD. Here we also show how specific initial states avoid the build up of heavy Pauli strings, and we observe specific combinations of observables and initial states where the dynamics appears to be dominated only by light strings. Such combinations should be easy to describe using TEBD.

As an inset on each panel, we include the temporal evolution of the OWE, which correctly reflects the complexity of simulating the dynamics of the chosen operator $\sigma^x$ through TEBD, and strongly depends on the initial state. As expected, for state $|0,~0\rangle$ the growth reaches higher values (close to the double) of those attained on state $|\frac{9\pi}{10},~0\rangle$.

    \subsection{Connection between the OWE and the equilibration temperature of different initial states}\label{ssec:OWE_temp}

In the previous section we have presented two extreme cases of local dynamics, one where the OWE is bounded versus another one where it would keep growing; given our finite computational resources we set $\omega^*=12$, forcing the OWE of the latter to saturate close to the upper bound for the entropy. As mentioned in former sections, the states with lower temperatures appear to require much more Pauli weight than those with higher temperatures, and thus one would be tempted to conclude that the different behavior of the OWE reflects such a temperature dependence, something that we address in detail in the remainder of this work. 

\begin{figure*}[!htb]
\includegraphics[width=1.\textwidth]{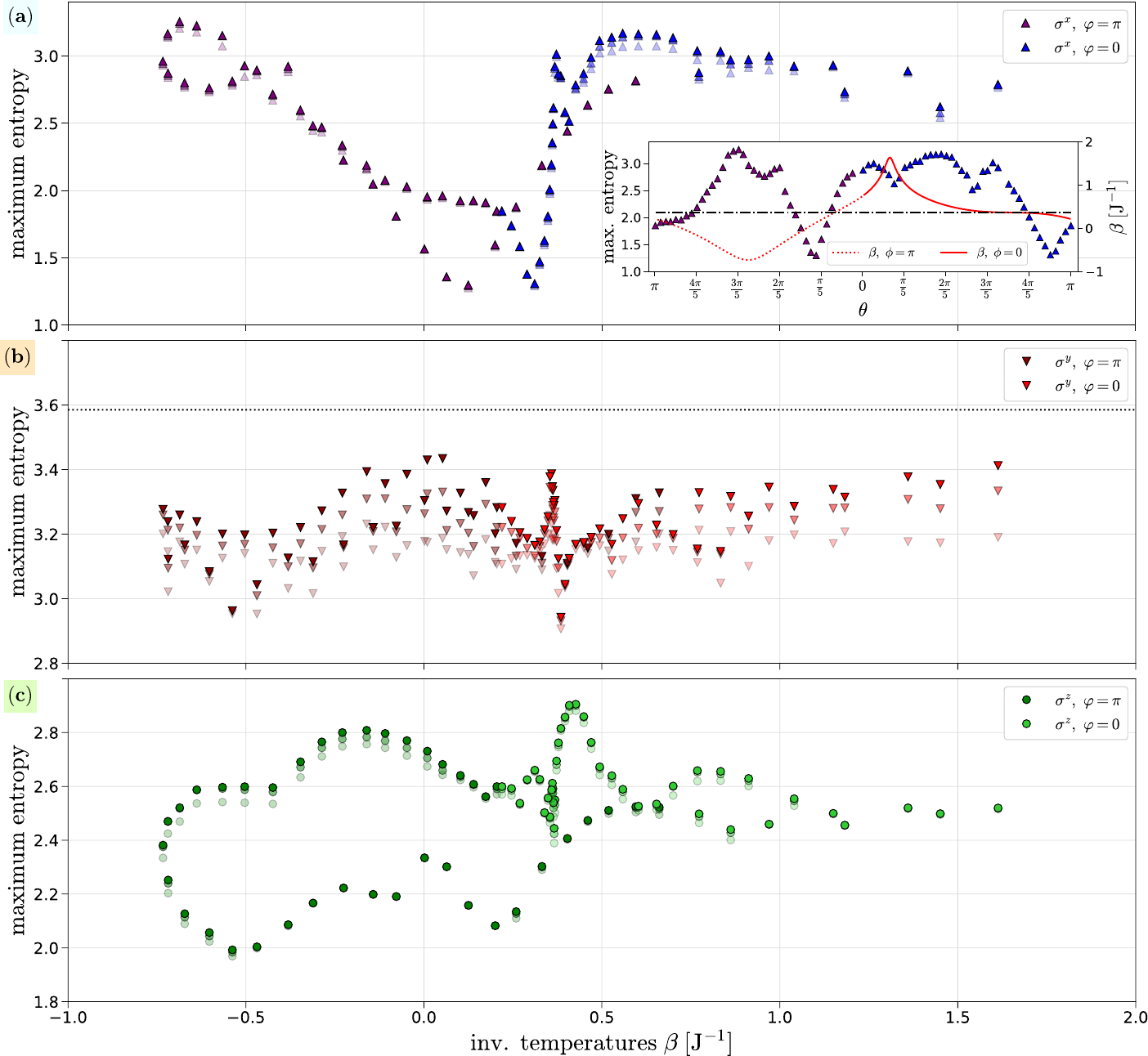}
  \caption{\textbf{Maximum OWE in the time interval $t\mathrm{J}\in[0,~ 5]$ for a set of local operators $\{\sigma^x,~\sigma^y,~\sigma^z\}$ and initial states on the sagittal cut of Bloch sphere defined by $\varphi=0$ (light color) and $\varphi=\pi$ (dark color)}. For each initial state, we show convergence for the entropies computed on consecutive maximum Pauli weights $\omega^*=10,~11,~12$, from most to least transparent respectively. {\bf{(a)}} The transverse magnetization $\sigma^x$ presents the highest variability upon changing the initial state, ranging from states with maximum OWE close to the cutoff set by computational resources $\log_2 12\simeq 3.58$, and two minima located in the inverse temperature interval $\beta\in[0,~0.5\mathrm{J}]$.  Despite an apparent tendency to higher entropies for bigger values of $|\beta|$, it is unclear whether this is a well-defined behaviour in the temperature, or whether there are other effects owed to the initial state. The same data profile is presented in terms of the polar angle as an inset, together with the temperature curve $\beta(\theta,~\varphi)$ in red, and the value of temperature at which the entropy appears to drop, $\beta^*\simeq 0.35$. {\bf{(b)}} Opposed to the former case, the $\sigma^y$ magnetization depends much less on the initial state and is close to saturating our computational resources, which consider a maximal $\omega^*=12$; despite that, each state is well-converged  in $\omega^*$. {\bf{(c)}} The longitudinal magnetization $\sigma^z$ displays an intermediate behavior clearly splitting in two branches of states corresponding to $\varphi=0$ and $\pi$ within the same range of temperatures.}
 	\label{fig:40states}
\end{figure*}

We repeat a similar analysis to the one in  Fig. \ref{fig:histograms} for a set of 40 states on the sagittal cut of the Bloch sphere defined by $\varphi=0$ and $\varphi=\pi$, and the three local operators $\{\sigma^x,~\sigma^y,~\sigma^z\}$. Instead of representing the full evolution of the OWE, we focus on the maximum it attains during the short time dynamics $t\mathrm{J}\in [0,~5]$. The results are then plotted as a function of the inverse temperature of the initial states in Fig. \ref{fig:40states}, and we differentiate the states on $\varphi=0$ from $\varphi=\pi$ with a two-color code on each panel. In order to reflect the effect of the approximations due to the computational Pauli weight threshold, we compute the maximum OWE using different thresholds $\omega^* =10,~11,~12$; the results are presented as markers with increasing opacity, being $\omega^*=10$ the faintest marker and $\omega^*=12$ the most opaque. In many cases, the three shades of the marker are not distinguishable (see e.g. panel (a) for $\beta \simeq 0.5$), hinting the fact that the maximum of the OWE for $\omega^*=10$ is a good approximation to its actual value. Such a convergence is seen for evolution where the OWE  stabilizes to small values compared to the numerical upper bound $\log_2 \omega^*$, that is $\log_2 12\simeq 3.58$.  In other cases we see big residual effects on the OWE induced by the cut-off $\omega^*$. There our predictions cannot be considered as converged.

In panel (a) we observe the maximum OWE for the x-magnetization attained within the studied time interval. This transverse x-magnetization is noticeably the observable with highest range of complexities compared to panels (b) and (c). Despite the states on each side of the Bloch sphere split into two branches for equal temperatures (blue for $\varphi=0$ and purple for $\varphi=\pi$), there seems to be a clear trend where the OWE is correlated with the equilibration temperature of the initial state. The expectation value of $\sigma_x$ in states equilibrating at the highest $\beta$ (lowest positive temperatures), require heavier Pauli strings to reach accurate values, with maximum OWE relatively close to the numerical upper bound. The state $|0,~0\rangle$, which has been studied in former sections, belongs to this regime.

As the equilibration temperature of the initial state rises ($\beta$ decreases), an  abrupt transition of the maximum OWE is observed around $\beta^*\simeq 0.35$, reaching a minimum close to 1 (regime in which we find  the previously analysed states $|\frac{\pi}{4},~\pi\rangle$ and $|\frac{9\pi}{10},~0\rangle$). The maximum OWE rises again as we move deep into the negative temperature regime. Overall, the results appear consistent with an increased complexity of the TEBD simulations in Heisenberg picture for low temperatures, suggesting a transition temperature between initial states that are easy and hard to simulate at $|\beta|\simeq 0.35$ in the plot.

As an inset, we present the same maximum OWE profile as a function of the polar angle on the Bloch sphere, (whose scale is on the left vertical axis) superimposed with the red curve representing the inverse equilibration temperature of the state (whose scale is on right vertical axis). This inset clarifies that the inverse temperature is not able to fully explain the behavior of the OWE. 

Panel (b) contains the maximum OWE for the y-magnetization, that opposed to the x-magnetization shows a very weak dependence on the initial state and its equilibration temperature. At the level of precision set by $\omega^*=12$, this plot is almost featureless given that our estimate of the OWE strongly depends on the cutoff for almost all states, and the values we obtained are all very close to the trust threshold 3.58. We expect that $\langle \sigma^y (t\to\infty)\rangle=0$ by the symmetry of the model, but our results seem to indicate that this 0 may appear as a subtle cancellation of contributions from many Pauli weight subspaces. We suspect that this behavior might be related with the absence of a direct overlap between $\sigma^y$ and the Hamiltonian, as elaborated in \cite{khemani2018operator}. Interestingly enough, despite the plot is almost featureless, the vertical drop around $\beta=0.35$ appears to be robust (the different shades of the markers tend to collapse there), coinciding with the location of the sharp transition observed in panel (a).

Finally, panel (c) contains the maximum OWE for the evolution of the z-magnetization, which explores intermediate values between those attained by $\sigma^y$ and the minima found in $\sigma^x$. In particular, this observable clearly shows the largest residual footprint of the initial state, as different states with the same equilibration $\beta$ display clearly different OWE. Here the special temperature $\beta^*$ at which a sudden change of the entropy occurs is only observable on the $\varphi=0$ branch.

The complex phenomenology we have just described is not completely unexpected given 
our limited computational resources, which severely restrict the range of times where we can compute the OWE. We thus cannot really observe the simplifications induced by the saturating $\mathrm{OWE}(t\to\infty)$ allowing to map the complexity of the evolution to an effective equilibrium temperature, and we  clearly observe a residual dependence on the initial states.  

However our results are of practical relevance, since they unveil the presence of an OWE barrier at relatively short times that clearly depends both the initial state and the operator to be simulated in the Heisenberg picture. As such, our findings suggest that some combinations of states and operators are easier to simulate than others. These pairs of operators and states provide instances of quantum dynamics that can be solved via classical simulations (e.g. by truncating the orthogonal Pauli weight during the evolution \cite{marimon2024decoherence}). 

Importantly, even in the vicinity of the easier combinations of $\hat{\mathcal{O}}$ and $\hat{\rho}(t=0)$ we find abrupt transitions where just slight changes in the state could generate OWE barriers that could be hard to circumvent with traditional techniques. 

\begin{figure}[!htb]
  \includegraphics[width=1.\columnwidth]{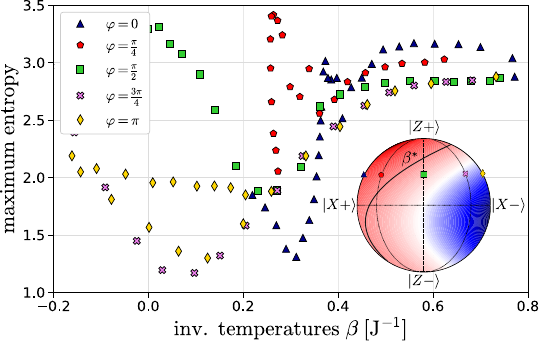}
  \caption{\textbf{Detail of the temperature interval where a transition in the maximum OWE appears to occur for $\langle\sigma^x (t)\rangle$}. For states placed across cuts defined by $\varphi = 0,~\frac{\pi}{4},~\frac{\pi}{2},~\frac{3\pi}{4},~\pi$ around the potential critical temperature $\beta^*\simeq 0.3$ we observe similar drops of the maximum OWE in $t\mathrm{J}\in[0,~\mathrm{5}]$. On top of the presence of entropy minima there appears to be another dependence on the Bloch sphere angles defining initial states. We observe that the minimal entropy cases not only depend on temperature but also on the initial state; this is a generic feature despite the suggesting presence of an abrupt drop of the entropy.}
 	\label{fig:transition}
\end{figure}

In order to illustrate that the aforementioned transition in the OWE is generic, we repeat the simulations of the x-magnetization in Fig. \ref{fig:40states} (a) for other vertical cuts on the Bloch sphere within the same interval of temperatures where we found a complexity drop. The results are presented in Fig. \ref{fig:transition}, where different families of states belonging to the new cuts are defined by $\varphi=\frac{\pi}{4}$ (red pentagons), $\varphi=\frac{\pi}{2}$ (green squares), $\varphi=\frac{3\pi}{4}$ (violet crosses) on top of the former cuts with $\varphi=0$ (blue triangles) and $\varphi=\pi$ (golden diamonds). Despite the clear deviations originating from different initial states with the same equilibration temperature, we identify drops on the complexity in all the cuts within $0.2\leq \beta \leq 0.4$, though with different slopes: $\varphi=0 $ and $\frac{\pi}{4}$ display abrupt jumps with strong slope while the rest of cuts follow a common, smoother slope.


\section{SUMMARY, OPEN QUESTIONS AND OUTLOOK.}\label{sec:V}

In this work, we have conducted a comprehensive analysis of the complexity underlying the simulation of the dynamics of local operators in the Heisenberg picture. Using the initial state of the quench allows distinguishing between the subspaces of evolved operators containing parallel and orthogonal Pauli strings. We have shown that the operator norm flows between these subspaces are increasingly suppressed as their Pauli weight grows. This insight allows one to focus predominantly on the parallel Pauli strings when addressing the complexity of encoding the matrix elements selected by the initial state of the time-evolved operators.

Our results demonstrate that in most cases, the dynamics quickly generates high-weight Pauli strings, exceeding the limits of current computational resources. The evolution of the Operator Weight Entropy (OWE) serves as a useful tool in tracking this phenomenon, growing logarithmically for operators that are difficult to simulate, and saturating for those that are computationally easier. By charting the maximal OWE over time, we have mapped the complexity landscape across different homogeneous product initial states on the Bloch sphere.

Notably, for a range of operators, we observe a robust signature of an abrupt change in complexity, linked to a narrow band in inverse temperature $\beta$. In the case of the $\sigma_x$ operator, this signature delineates a transition between high- and low-complexity regimes. While other operators predominantly exhibit high-complexity dynamics, the footprint of this transition remains evident. Whether this intermediate-time complexity transition reflects a transition in complexity of the late-time equilibrium states warrants further investigation.

Independently of this narrow $\beta$ line, our analysis provides a valuable framework for identifying combinations of operators and initial states that remain tractable within current computational limits. This sets clear boundaries for the potential scenarios of quantum computational advantage and offers a benchmark for the development of improved TN algorithms to address the more challenging scenarios. In future work we will focus on refining these TN methods using the insight from the present study, which represents a critical step towards a deeper understanding of the complexity in simulating the out-of-equilibrium dynamics of many-body quantum systems.

\section{Acknowledgements.}

We acknowledge valuable discussions on this and related
subjects with M.-C. Bañuls, A. Franco, P. Kos, G. Styliaris. C.R.M. acknowledges support from a “la Caixa” Foundation fellowship
(ID No. 100010434, code LCF/BQ/DI21/11860031).
LT  acknowledges support from the Proyecto Sinérgico CAM Y2020/TCS-6545 NanoQuCo-CM,
the CSIC Research Platform on Quantum Technologies PTI-001, and from the Grant TED2021-130552B-C22 funded by MCIN/AEI/10.13039/501100011033 and by the ``European Union NextGenerationEU/PRTR'',
and Grant PID2021-127968NB-I00 funded by MCIN/AEI/10.13039/501100011033.

\appendix
\section{Details on the MPO implementations.}\label{app:MPOs}

In this appendix we specify the particular MPOs that have been used to split the Heisenberg picture operators into contributing and non-contributing superpositions, and the projectors onto subspaces with defined Pauli weight.

In order to extract the contributing part with TNs, we define the projector onto strings with only parallel insertions as
\begin{equation}
\begin{split}
     &|\hat{\mathcal{O}}^{\mathrm{c}}(t)\rangle = \hat{\mathcal{P}}^{\mathrm{c}}|\hat{\mathcal{O}}(t)\rangle \quad\mathrm{with} \\ \hat{\mathcal{P}}^{\mathrm{c}}=\bigotimes_{j=1}^{L}&\big(|\mathds{1}_{j}\rangle\langle\mathds{1}_{j}|+|\sigma^{\parallel}_{j}\rangle\langle\sigma^{\parallel}_{j}|\big)=\bigotimes_{j=1}^{L}W_{\mathrm{c}},
\end{split}
\end{equation}
\noindent which is an MPO with trivial bond dimension 1. For a comprehensive discussion on the basics of MPOs and their construction, we refer the reader to Sec. 5 of \cite{schollwock2011densitymatrix}.

The orthogonal part of the operator may include any number of parallel insertions, but it will always include at least one orthogonal insertion ($\omega^{\perp}\geq 1$) in some site; in this way, the projector onto the non-contributing subspace $\hat{\mathcal{P}}^{\mathrm{nc}}$ can be cast as an MPO specified by the transfer tensor
\begin{equation}
    W_{\mathrm{nc}} = \begin{pmatrix}
        W_{\mathrm{c}} & \sum_{i=1,2} |\sigma^{\perp,i}\rangle\langle \sigma^{\perp,i} | \\
        0 & \mathds{1}_4 
    \end{pmatrix}
\end{equation}
\noindent which passes from its initial internal state to its end state by detecting an orthogonal Pauli matrix. The boundaries are given by vectors with components $(W_{\mathrm{nc, L}})^{\alpha}=(W_{\mathrm{nc}})^{0, \alpha}$ and $(W_{\mathrm{nc,R}})^{\alpha}=(W_{\mathrm{nc}})^{\alpha, 1}$ for $\alpha=1,2$.

The splitting of each superposition into linear combinations of strings with a fixed Pauli weight is implemented by a \textit{Pauli weight projector}
$\hat{\mathcal{P}}_{\omega}$, which takes the form of an MPO with transfer tensor of rank $\omega + 1$:
\begin{equation}
    W_{\omega} =\begin{pmatrix}
        |\mathds{1}\rangle\langle\mathds{1}| & \Sigma & 0 & \cdots & 0 \\
        0 & |\mathds{1}\rangle\langle\mathds{1}| & \Sigma & \cdots & 0 \\
        \vdots &  & \ddots & & \vdots \\
         &  &  &  & \Sigma\\
        0 & & \cdots & & |\mathds{1}\rangle\langle\mathds{1}|
    \end{pmatrix}
    \label{eq:transf_matr_w_projector}
\end{equation}
\noindent where we define the sum of projectors onto particular Pauli matrices $\Sigma =\sum_\alpha |\sigma^\alpha\rangle\langle \sigma^\alpha|$. The boundaries are once again set by $(W_{\omega,\mathrm{L}})^{\alpha}=(W_{\omega})^{0, \alpha}$ and $(W_{\omega,\mathrm{R}})^{\alpha}=(W_{\omega})^{\alpha, 1}$ for $\alpha=1,2$.

Note that one can construct projectors onto subspaces with only fully parallel or fully orthogonal strings by combinations of the former projectors, or by substituting $\Sigma$ in Eq. \ref{eq:transf_matr_w_projector} by the sum of projectors onto only parallel or only orthogonal Pauli matrices.

\bibliographystyle{apsrev4-2}
\bibliography{my_ref.bib}

\end{document}